\documentclass{PoS}
\usepackage{amssymb,amsmath,graphicx}

\title{Probing novel TeV physics through precision calculations of scalar and tensor charges of the nucleon}
\ShortTitle{Probing novel interactions at TeV scale in neutron decays}

\author{\speaker{Rajan Gupta,} \footnote{LA-UR-12-26832} \ \ Tanmoy Bhattacharya and Anosh Joseph 
\\
Theoretical Division, Los Alamos National Laboratory, Los Alamos, NM 87545, USA \\}
\author{Saul D. Cohen and Huey-Wen Lin \\
Department of Physics, University of Washington, Seattle, WA 98195\\}

\abstract{We present an update on the calculation of matrix elements
  of iso-vector scalar, axial and tensor charges between a neutron and
  a proton state.  These matrix elements are needed to probe novel scalar
  and tensor interactions in neutron beta-decay that can arise in
  extensions of the Standard Model at the TeV scale.  Our calculations
  are being done using valence clover fermions on dynamical
  $N_f=2+1+1$ HISQ configurations generated by the MILC
  Collaboration. We provide preliminary estimates of the dependence of
  these matrix elements on the light quark masses, lattice spacing,
  and the time separation between the source and sink of the nucleons.
  We also find that the renormalization constants calculated using the
  RI-sMOM scheme are close to unity for the HYP smeared HISQ lattices. }

\FullConference{The 30 International Symposium on Lattice Field Theory - Lattice 2012,\\
		June 24-29, 2012\\
		Cairns, Australia}

\begin{document}


\section{Introduction}

The observed electroweak symmetry breaking in the Standard Model (SM)
points to new physics at the TeV scale, which is currently being
explored at the LHC.  There are many candidate extensions of the SM,
such as supersymmetry and extra dimensions, but, so far, there is no
experimental guidance on what new interactions and particles exist at
this scale. A theoretical approach (effective field theory) is to
postulate new interactions at the TeV scale (without necessarily
specifying a model) and analyze the corrections they would give rise to
in low-energy (few-GeV) physics. Precision measurements of decays can
then be used to bound the allowed parameter space of the posulated
interactions. In Ref.~\cite{Bhattacharya:2011qm}, we have shown that precision
measurements of decays of (ultra)cold neutrons at the $10^{-3}$ level
can be used to improve the bounds on new scalar and tensor
interactions at the TeV scale provided estimates accurate to 10--20\%
of the matrix elements of isovector scalar and tensor bilinear quark
operators are realized.  

In these proceedings, we will summarize the progress we have made in
the needed calculations to extract the charges $g_A$, $g_S$ and $g_T$.
These charges are defined as
$$
g_A = Z_A  \langle p | O_A | n \rangle \ ; \ \ \ 
g_S = Z_S  \langle p | O_S | n \rangle \ ; \ \ \
g_T = Z_T  \langle p | O_T | n \rangle \ ,
$$ where $Z_\Gamma$ are the corresponding renormalization constants,
which we calculate nonperturbatively in the RI-sMOM
scheme~\cite{RI_sMOM:rome,RI_sMOM:bnl}.  This report is a follow up to
Ref.~\cite{Bhattacharya:2011qm}, and we will use the notation
established there. The calculations are being done using the 2+1+1
flavor HISQ lattices being generated by the MILC
Collaboration~\cite{MILC}. Clover fermions on HYP smeared HISQ
lattices are being used for constructing the correlation functions.
Three values of the lattice spacing are being analyzed, $a=0.12$,
$0.09$ and $0.06$ fm, to perform the continuum extrapolation, and at
each lattice spacing we analyze two values of the light quark masses
corresponding to $M_\pi \approx 310$ and $220$ MeV.  We will address
the following sources of statistical and systematic errors in the
extraction of matrix elements and, from them, the charges.
\begin{itemize}
\item
The signal and the statistical errors in the two- and three-point
correlation functions. 
\item
The dependence of $g_{A,S,T}$ on the light-quark (pion) masses and
lattice spacing to estimate uncertainty due to  
chiral and continuum extrapolations.
\item
The effect of contamination by excited states. 
\item
Estimates of the renormalization constants calculated 
in the RI-sMOM scheme.
\end{itemize}
A similar analysis has been reported in Ref.~\cite{Jeremy} at this conference.

\section{Statistics}

The MILC Collaboration has produced ensembles of roughly 5500
trajectories of 2+1+1 flavor HISQ lattices at each of the six values
of quark masses and lattice spacings we are analyzing as described
in~Table~\ref{tab:statistics}.  Five hundred trajectories are discarded for
thermalization.  Configurations are then analyzed separated by five
trajectories.  On each configuration, four smeared sources, displaced
both in time and space directions to reduce correlations, are
used. Furthermore, two sets of these four source points, again
maximally separated in space and time directions, are used on each
alternate configuration to reduce correlations.  The roughly 500
configurations with each of these two sets of sources are also analyzed
separately. We verify that the two sets give compatible results and the errors are roughly $\sqrt 2$
larger compared to the full set. Our main conclusion is that with 1000 or
more configurations one gets a statistically significant signal needed
for obtaining the design accuracy of 10--20\% in both the 2-point and
3-point functions for all cases.  Extraction of $g_S$ is the nosiest and 
drives the overall sensitivity as discussed below.

\begin{table}[thbp]
\begin{tabular}{|c|c|c|c|c|c|c|c|}
\hline
$a$ (fm) & $m_l/m_s$  &  $M_\pi$ (MeV) &  $L^3 \times T$  & $M_\pi L$ & Configs (Z) & Configs (ME) & $\Delta t$  \\
\hline
0.12 & 0.2   & 305  & $24^3 \times 64$   & 4.54   &  50 & 1013/1013 & 8,9,10,11,12 \\
0.12 & 0.1   & 217  & $32^3 \times 64$   & 4.29   &  50 & 958/958   & 8,10,12 \\
\hline
0.09 & 0.2   & 313  & $32^3 \times 96$   & 4.5    &     & 391/1000  & 12,14 \\
0.09 & 0.1   & 220  & $48^3 \times 96$   & 4.73   &     & 443/1000  & 12,14 \\
\hline
0.06 & 0.2   & 320  & $48^3 \times 144$  & 4.53   &     & 330/1000  & 16,20 \\
0.06 & 0.1   & 229  & $64^3 \times 144$  & 4.28   &     & 0/1000    &  \\
\hline
\end{tabular}
\caption{Lattice parameters of the six ensembles of 2+1+1 flavor HISQ
  lattices analyzed. We give a status report (as of November 2012) on
  the number of analyzed and total configurations used in calculations
  of ME and for the calculation of renormalization constants (Z). The last 
  column gives in lattice units the time seperations $\Delta t$ between source and sink 
  investigated to understand the excited-state contamination. }
\label{tab:statistics}
\end{table}

\section{Excited-State Contamination}

The desired matrix elements need to be calculated between ground state
nucleons. The operators used to create and annihilate the states,
however, couple to the nucleon and all its excited states.  There are
two possible ways to reduce contribution from excited states: by
reducing the overlap of the interpolating operator with the excited
states and by increasing the time separation $\Delta t$ between the source
and sink to exponentially suppress excited-state contamination.
Statistics limit the upper value of $\Delta t$  that can be explored. We use
smeared sources and sinks to reduce overlap and investigate up to five
time separations.

\begin{figure}
\begin{tabular}{cc}
\includegraphics[trim=0 0 153 204,clip=true,width=.48\textwidth]{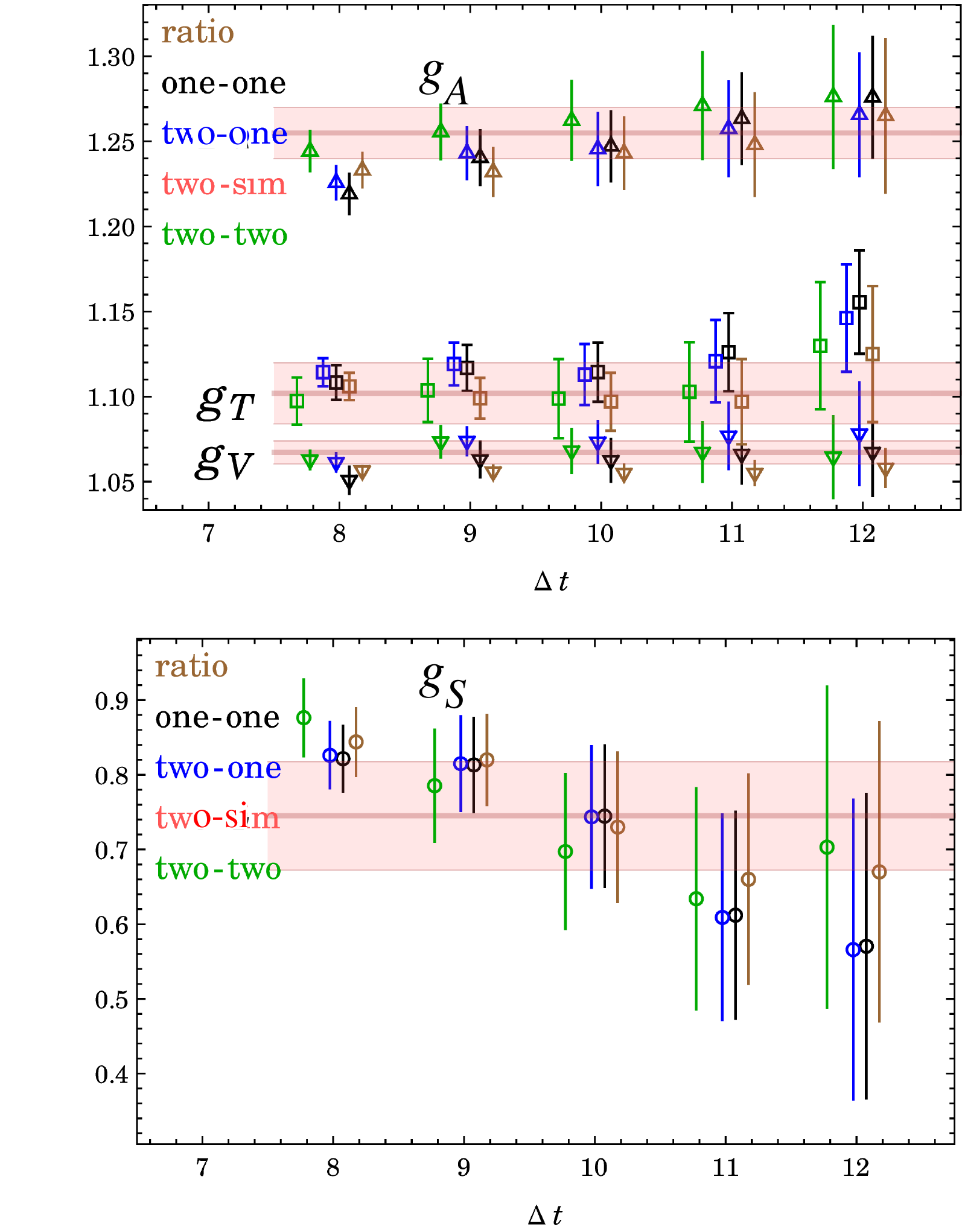} & 
\includegraphics[trim=0 0 153 204,clip=true,width=.48\textwidth]{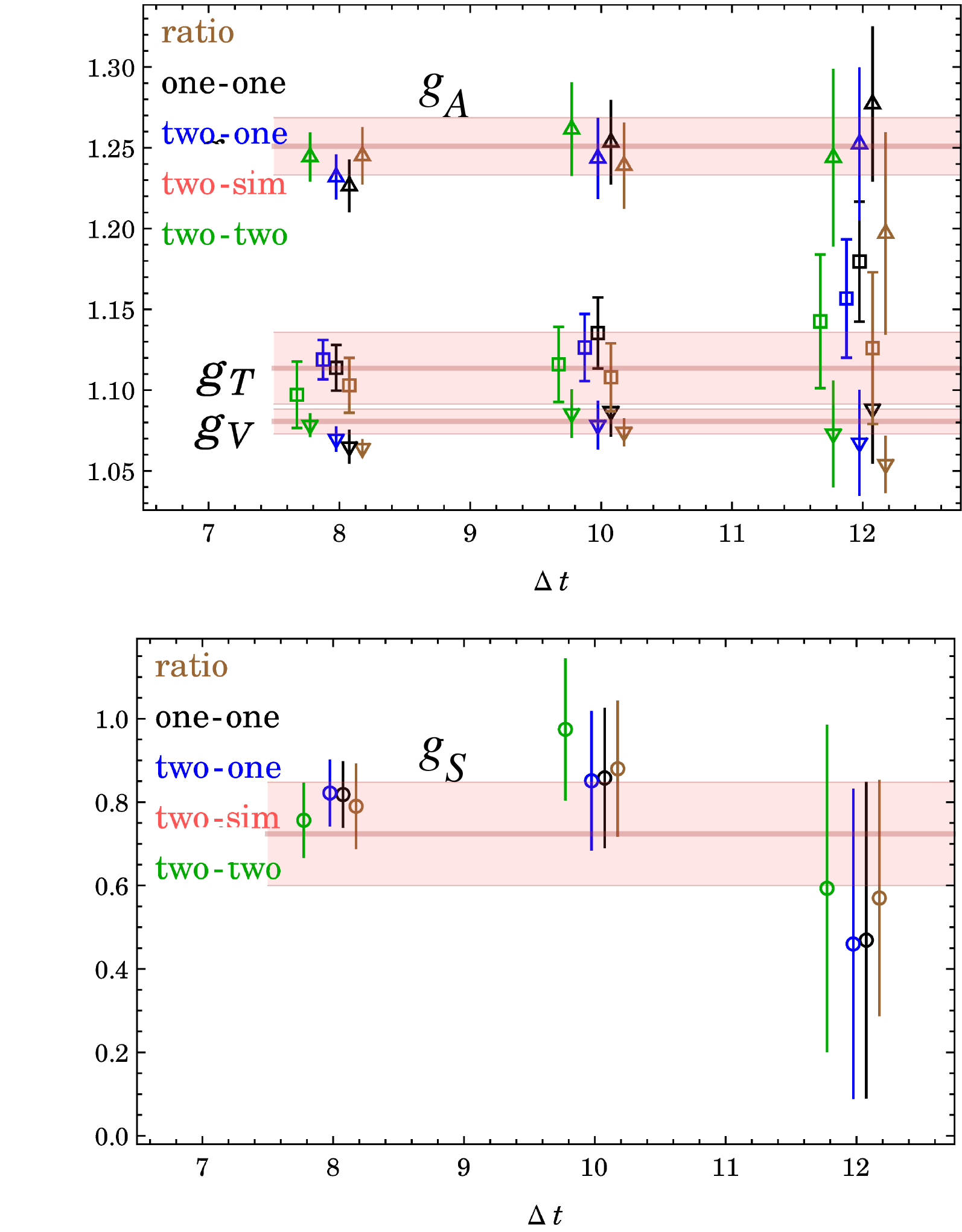} \\
\end{tabular}
\caption{
\label{Fig:G}
Estimates of the unrenormalized charges $g_A$, $g_V$, $g_S$ and $g_T$
from the five different fits for the $a=0.12$ fm lattices with $M_\pi$
= 310 MeV (left) and 220 MeV (right). The bands correspond to the
{\it 2-2 simultaneous fit} described in the text and are used as our best estimates. The other
four estimates are consistent with this estimate. 
$\Delta t$ is the separation in time between the source and the sink 
in lattice units.}
\end{figure}

Assuming that all excited-state contamination can be represented by a
single excited state with amplitude $A_1$ and mass $M_1$, we can write
the 3-point function with source at $t=0$, operator insertion at
$t=t$ and sink at $t= \Delta t$ as
\begin{eqnarray}
{\cal C}_\Gamma
  &=& |A_0|^2 \langle 0 | O_\Gamma | 0 \rangle  e^{-M_0 {\Delta t}} + 
      |A_1|^2 \langle 1 | O_\Gamma | 1 \rangle  e^{-M_1 {\Delta t}} + 
\nonumber\\
  & & A_0A_1^* \langle 0 | O_\Gamma | 1 \rangle  e^{-M_0 t} e^{-M_1 ({\Delta t}-t)} + 
      A_0^*A_1 \langle 1 | O_\Gamma | 0 \rangle  e^{-M_1 t} e^{-M_0 ({\Delta t}-t)} 
\label{eq:three-pt}
\end{eqnarray}
where $\Gamma$ represents one of the sixteen Clifford elements
defining the bilinears.  To extract $\langle 0 | O_\Gamma | 0
\rangle$ from the 2- and 3-point functions we examine five different
fits. All errors are estimated with the full analysis done within a
single-elimination jackknife method. 
\begin{itemize}
\item
{\it 1-1 method} assumes a single state dominates the 2-pt and 3-pt
functions. $A_0$ and $M_0$ are extracted from a fit to the 2-pt
function and $\langle 0 | O_\Gamma | 0 \rangle$ are estimated from 
the 3-pt functions keeping only the first term in Eq.~(\ref{eq:three-pt}).
\item
{\it Ratio method} also assumes a single state dominates the 3-pt
function. $\langle 0 | O_\Gamma | 0 \rangle$ are estimated from the
ratio of 3-pt to 2-pt functions.  While some of the systematics cancel
in the ratio, this fit relies on there being a good signal in the 2-pt
function at separation $\Delta t$.
\item
{\it 2-1 method}: $A_0$, $A_1$, $M_0$ and $M_1$ are extracted from a
fit to the 2-pt function. Of these, only $A_0$ and $M_0$ are used to
estimate $\langle 0 | O_\Gamma | 0 \rangle$ by fitting to the first
term in Eq.~(\ref{eq:three-pt}).
\item
{\it 2-2 method} extracts $A_0$, $A_1$, $M_0$ and $M_1$ from a fit to
the 2-pt function. These amplitudes and masses are used in a
2-parameter fit to the 3-pt function to estimate $\langle 0 | O_\Gamma
| 0 \rangle$ and $\langle 1 | O_\Gamma | 0 \rangle$. We assume
$\langle 0 | O_\Gamma | 1 \rangle$ and $\langle 1 | O_\Gamma | 0
\rangle$ are equal and we analyze only the real part of the 3-point
function. We also neglect $\langle 1 | O_\Gamma | 1 \rangle$.
\item
{\it 2-2 simultaneous fit to all $\Delta t$} is the same as the {\it 2-2
  method} for extracting $A_0$, $A_1$, $M_0$ and $M_1$. The fit to the
3-point function uses data for all investigated values of $\Delta t$
simultaneously.  
\end{itemize}

Our current understanding of excited-state contamination is based on
the analysis of the full set of 1013 configurations for the $a=0.12$
fm $M_\pi=310$ MeV ensemble using $\Delta t=8,\ 9,\ 10,\ 11,\ 12 $ and 958
configurations for the $a=0.12$ fm $M_\pi=310$ MeV ensemble using
$\Delta t=8,\ 10,\ 12 $.  These data are shown in Fig.~\ref{Fig:G}. The
vector charge $g_V$ is shown only as a check. 

We find that the ratio method exhibits plateaus for $g_A$, $g_V$ and
$g_T$, so it is not surprising that the first four methods give
consistent estimates. The estimates from the {\it 2-2 simultaneous
  fit} method are shown by the horizontal bands which we take as our
best estimates. Our observations, based on these two ensembles of
roughly 1000 configurations and smeared sources used by us in
calculating quark propagators, are:

\begin{itemize}
\item
On the $0.12$ fm lattices, the statistical errors increase by about
$40$\% with each unit increase in $\Delta t$. We find a similar
increase for the $a=0.09$ and $0.06$ fm lattices once the unit
increase in $\Delta t$ is scaled by the factors $1.33$ and $2$ to keep
distances constant in physical units.
\item
There is a small trend showing an increase (within $1 \sigma$) in $g_A$ and $g_T$ with $\Delta t$. 
\item
For $t > 12$, even the signal in the 2-point nucleon correlator
becomes noisy. Based on the trends seen in the 
$M_\pi=310$ MeV ensemble, we considered it sufficient to 
investigate the $M_\pi=220$ MeV ensemble using $\Delta t =8,\ 10,\ 12$. 
\item
The {\it 2-2 simultaneous fit} estimates of the central values and errors
are consistent with data from other fits for all values of $\Delta t$. 
\item
The errors increase by about 20\% on lowering the light ($u$ and $d$) quark masses 
by a factor of two $i.e.$ from $M_\pi=310$ to $220$ MeV ensembles. 
\item
The signal in $g_S$ is the noisiest. On the $M_\pi=220$ MeV ensembles, the error 
estimate is $\approx 20\%$, reasonably close to the desired accuracy. 
\item
The statistical signal improves significantly as $a$ is reduced. This is presumably 
due to the larger lattice volumes and smoother gauge configurations. Preliminary 
results show that on the $a=0.09$ and $0.06$ fm ensembles we will be able to 
extract even $g_S$ within $10\%$ accuracy. 
\end{itemize}

Our conclusion, based on the $a=0.12$ fm lattices, is that the central values
from the {\it 2-2 simultaneous fit} agree with those from the other fits
for separation $\Delta t=10$ which corresponds to $\approx 1.2$~fm. We, therefore,
consider the $\Delta t=10$ ($i.e.$ in physical units this $\Delta t=1.2$~fm) data the best
compromise between reducing excited-state contamination and having a
good statistical signal. The choice of $\Delta t$ values investigated on the
$a=0.09$ and $0.06$ fm ensembles is based on these conclusions.

\begin{figure}[p]
\begin{tabular}{cc}
\includegraphics[width=.48\textwidth]{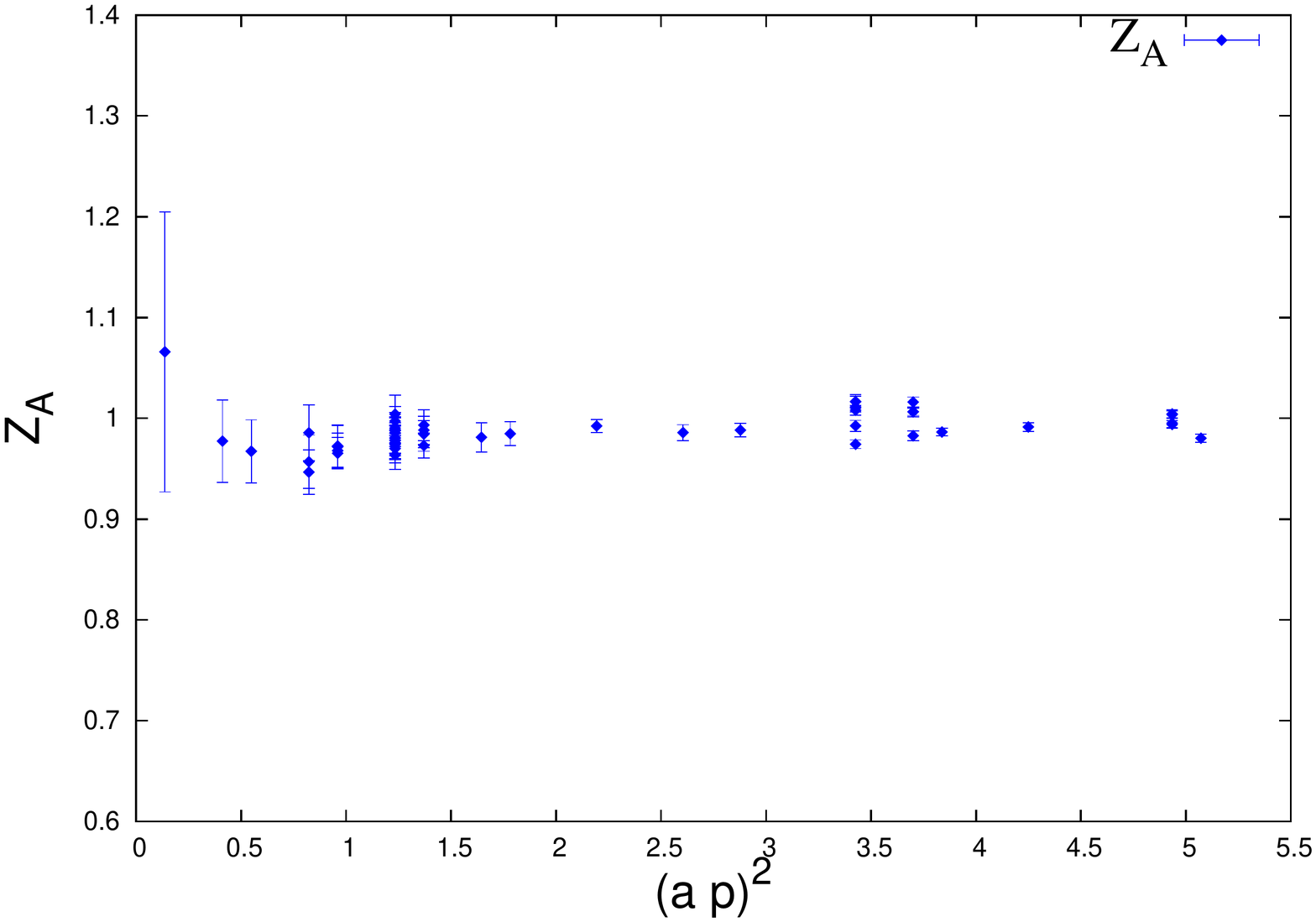} & \includegraphics[width=.48\textwidth]{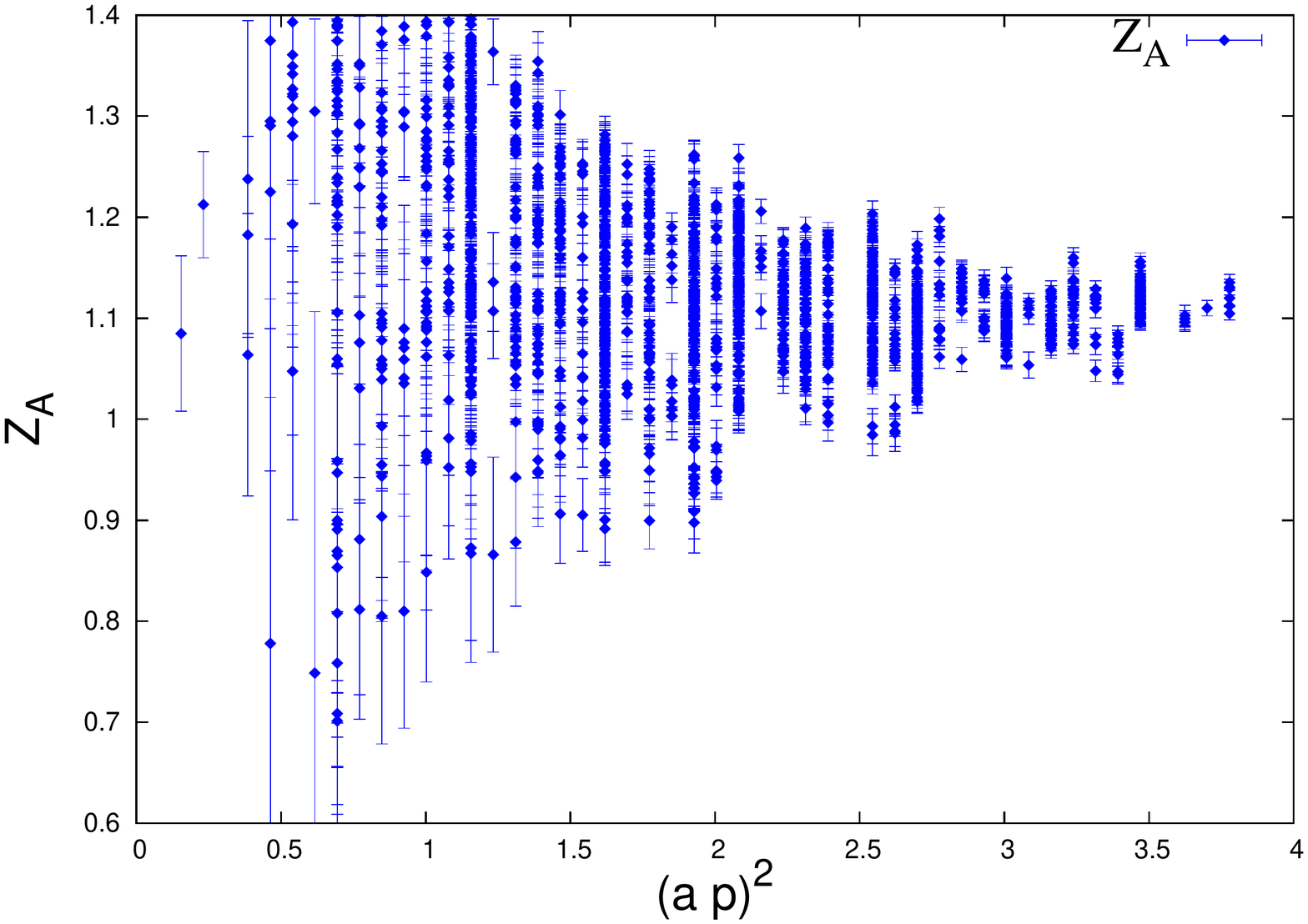} \\[-16pt]
$Z_A$ ($M_\pi=310$ MeV ensemble) &  $Z_A$ ($M_\pi=220$ MeV ensemble) \\
\includegraphics[width=.48\textwidth]{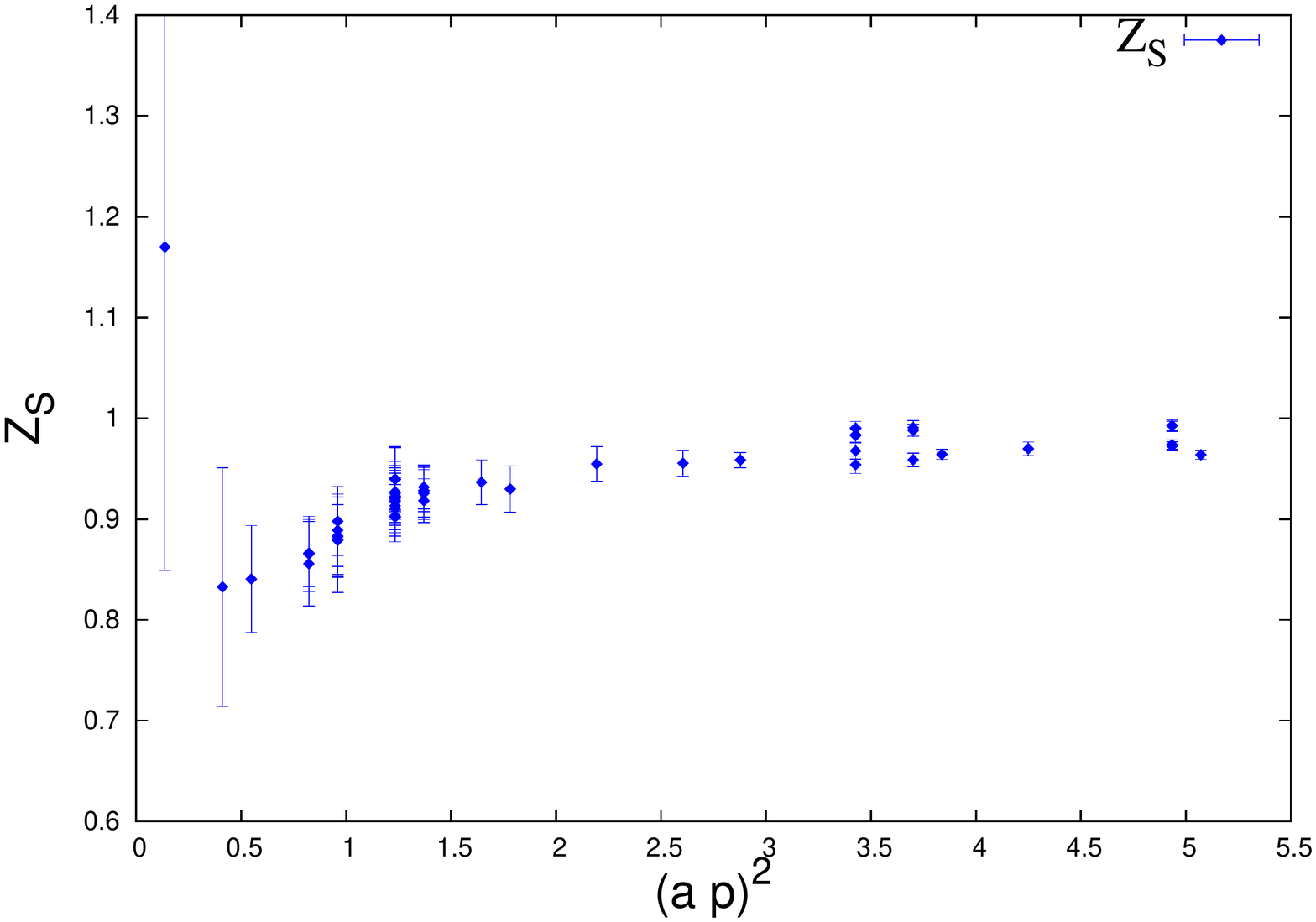} & \includegraphics[width=.48\textwidth]{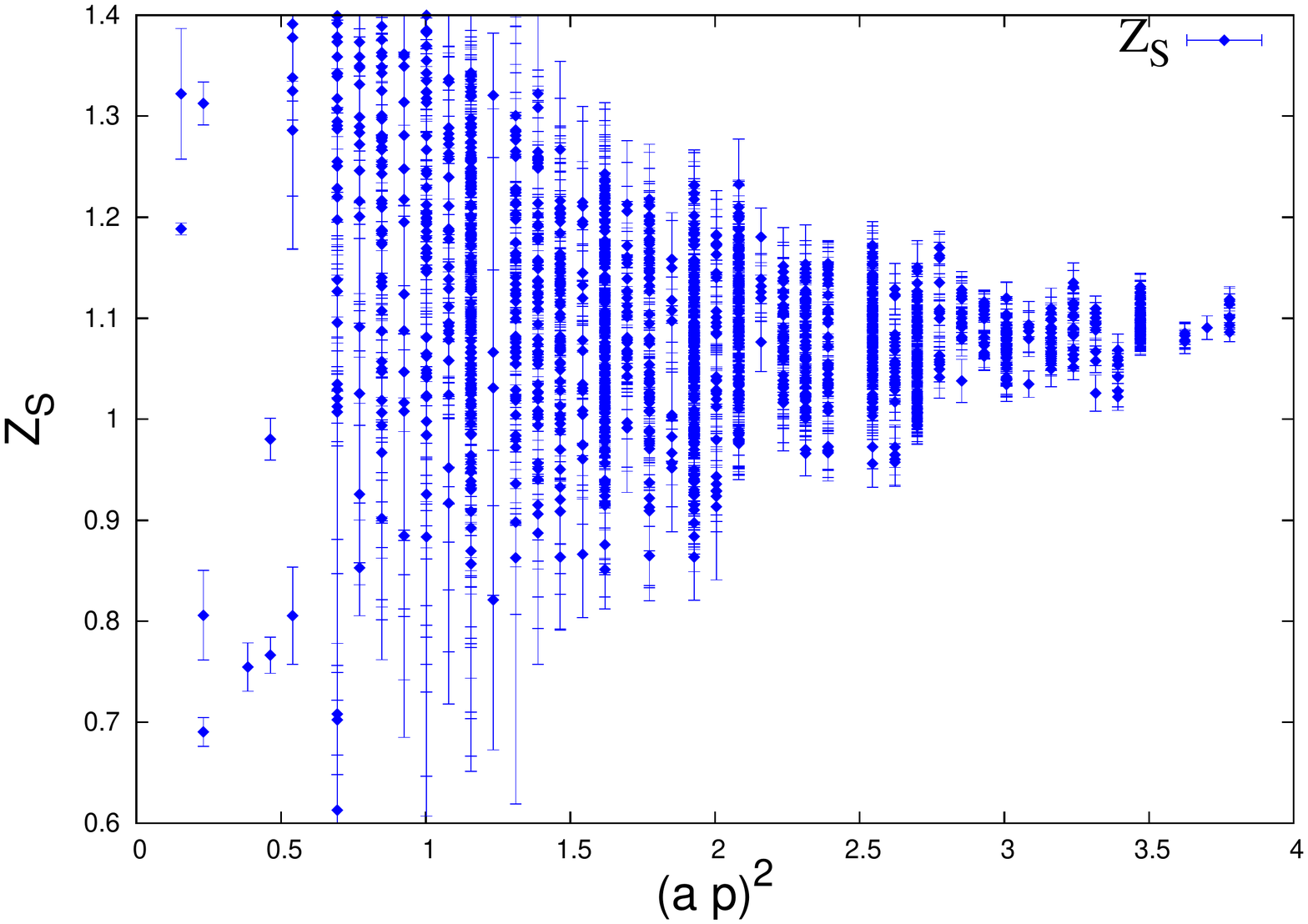} \\[-16pt]
$Z_S$ ($M_\pi=310$ MeV ensemble) &  $Z_S$ ($M_\pi=220$ MeV ensemble) \\
\includegraphics[width=.48\textwidth]{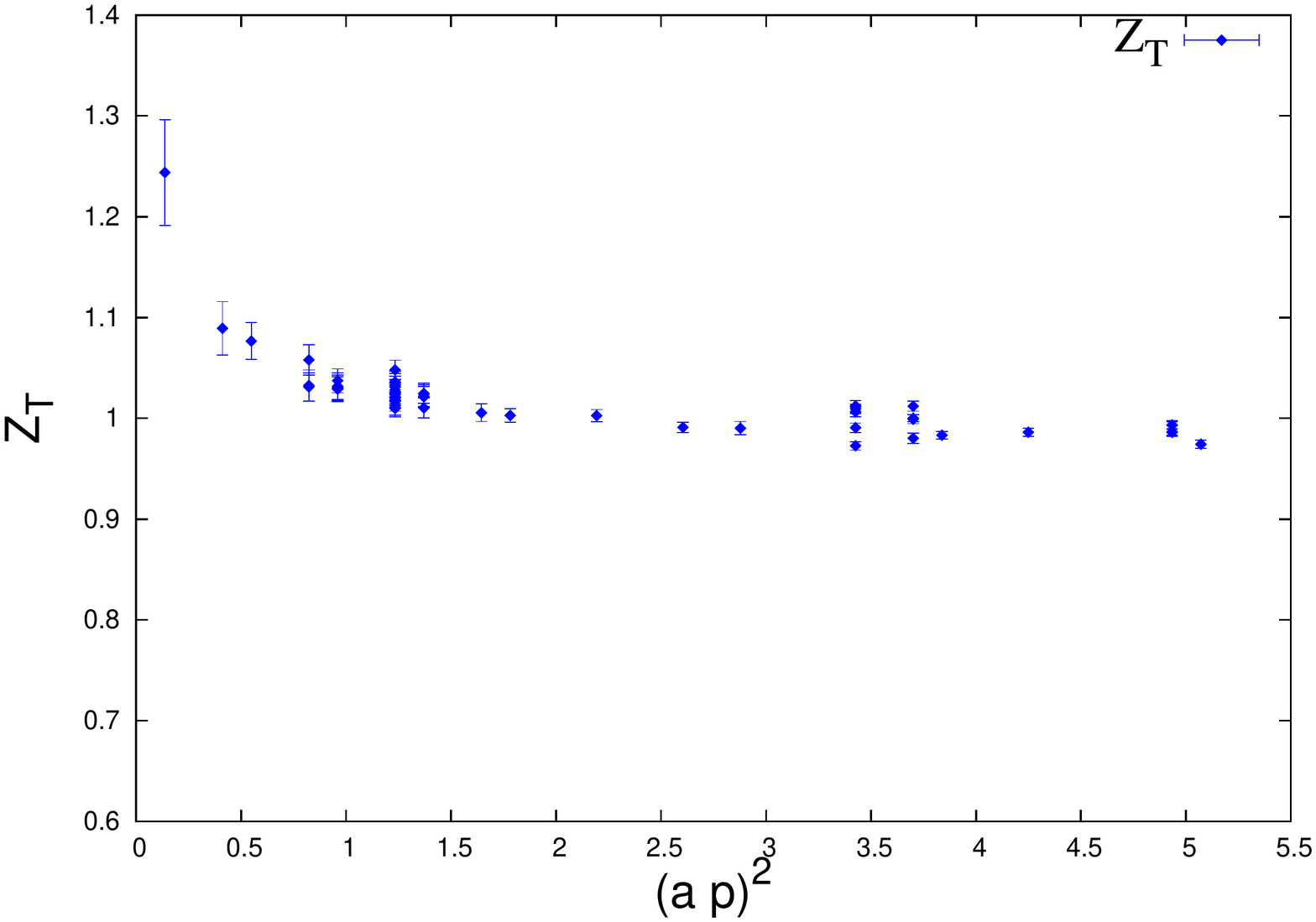} & \includegraphics[width=.48\textwidth]{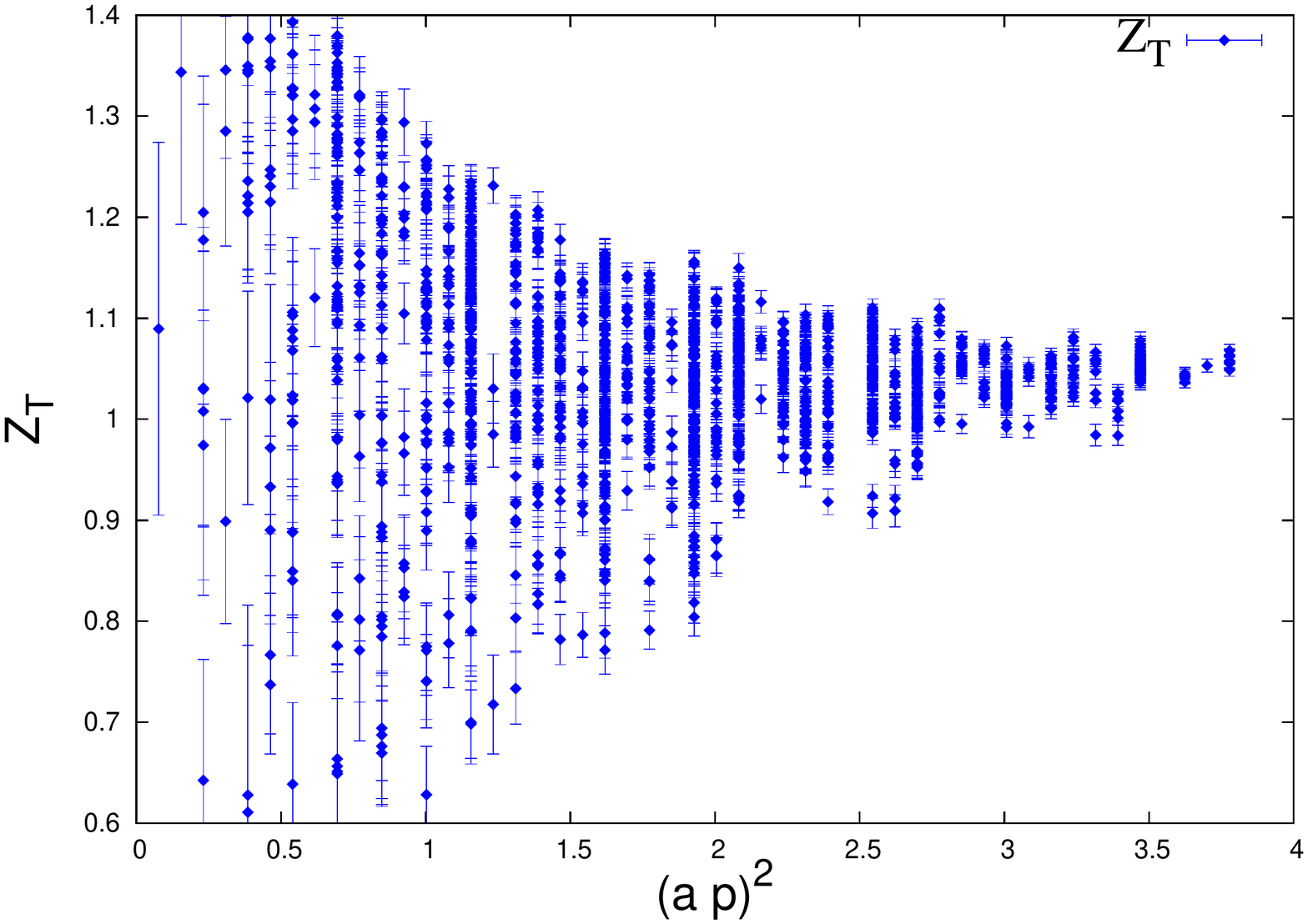} \\[-16pt]
$Z_T$ ($M_\pi=310$ MeV ensemble) &  $Z_T$ ($M_\pi=220$ MeV ensemble) \\[12pt]
\end{tabular}
\caption{
\label{Fig:Z}
The data for the renormalization constants $Z_A$, $Z_S$ and $Z_T$ for
the $a=0.12$ fm lattices as a function of the momenta $p_i^2 = p_f^2 =
(p_i-p_f)^2$.  Note that $p=1.6$~GeV for $pa=1$.  The analysis is done
using point source Wilson-Clover action propagators calculated on HYP
smeared HISQ lattices.  For these ensambles, we find that estimates of
all three $Z_\Gamma$ are within $10\%$ of unity. The dependence on the
light quark mass, $i.e.$, the $\sim$ 10\% increase on going from the
$M_\pi=310$ to $220$ MeV ensembles, is being studied.}
\end{figure}

\section{Dependence on Quark Mass and Lattice Spacing}

The data show a small increase in the values of the bare charges
$g_A$, $g_S$ and $g_T$ on going from $M_\pi=310$ to $220$ MeV ensemble
for $a=0.12$fm.  The change is, however, within $1\sigma$.  The data
for the $a=0.09$ and $0.06$ fm lattices are too preliminary to draw
conclusions on the dependence on either the quark mass or on the
lattice spacing. Since the statistical signal improves as $a \to 0$,
we anticipate that our final results will shed light on possible dependence
on the quark mass and the lattice spacing.

\section{Renormalization Constants $Z_\Gamma$}

The renormalization constants $Z_\Gamma$ are calculated in the RI-sMOM scheme
using point source propagators evaluated on Landau gauge fixed
lattices. The data for $a=0.12$ fm ensembles are shown in
Fig.~\ref{Fig:Z}. In this scheme only momenta satisfying $p_i^2$ =
$p_f^2$ = $(p_i-p_f)^2$ are examined (the $M_\pi=310$ MeV lattices have
much fewer momenta satisfying this condition due to the smaller
lattice volume), and the data have been averaged over points
equivalent under cubic symmetry. We find that for the HYP smearing
used in the calculation, all the renormalization constants are already
within $10\%$ of unity at $a=0.12$~fm. We are currently investigating
$O(p^4)$ contributions in these estimates to reduce these discretization effects. 

\section{Conclusions}

We show that the excited-state contamination is smaller than
statistical errors for source-sink separation greater than
$1.2$~fm. 
The renomalization constants in the RI-sMOM scheme are within
10\% of unity already on the coarsest lattices at $a=0.12$~fm. We
therefore, conclude that $O(1000)$ lattices will be sufficient to get
estimates with 10--15\% precision for both $g_S$ and $g_T$ at all
three values of the lattice spacing. We anticipate completing the
analysis on the $a=0.09$ and $0.06$ fm lattices over the next year.
These data are needed to confirm the small observed dependence on the
quark mass and lattice spacing, and before conclusions on the 
chiral and continuum extrapolations can be drawn. 

\section*{Acknowledgments}
We thank V. Cirigliano, M. Graesser and our experimental
colleagues for discussions. We thank the MILC Collaboration for sharing the HISQ
lattices.  These calculations were performed using the Chroma software
suite~\cite{Edwards:2004sx}.  Numerical simulations were carried out
in part on facilities of the USQCD Collaboration, which are funded by
the Office of Science of the U.S. Department of Energy, and the
Extreme Science and Engineering Discovery Environment (XSEDE), which
is supported by National Science Foundation grant number OCI-1053575.
The speaker is supported by the DOE grant DE-KA-1401020.

\end{document}